\documentclass[prb,aps,twocolumn,groupedaddress]{revtex4}

\usepackage{bm}
\usepackage{graphicx}
\usepackage{color}
\usepackage{ulem}

\begin{document}

\title{Nano-mosaic of Topological Dirac States on the Surface of Pb$_5$Bi$_{24}$Se$_{41}$ Observed by Nano-ARPES}

\author{Kosuke Nakayama$^1$, Seigo Souma$^{2,3}$, Chi Xuan Trang$^1$, Daichi Takane$^1$, Chaoyu Chen$^4$, Jose Avila$^4$, Takashi Takahashi$^{1,2,3}$, Satoshi Sasaki$^5$, Kouji Segawa$^6$, Maria Carmen Asensio$^7$, Yoichi Ando$^8$, and Takafumi Sato$^{1,2,3}$}

\affiliation{$^1$Department of Physics, Tohoku University, Sendai 980-8578, Japan\\
$^2$Center for Spintronics Research Network, Tohoku University, Sendai 980-8577, Japan\\
$^3$WPI Research Center, Advanced Institute for Materials Research, Tohoku University, Sendai 980-8577, Japan\\
$^4$Synchrotron SOLEIL, L'Orme des Merisiers, Saint-Aubin, BP 48, 91192 Gif-sur-Yvette C\'edex, France\\
$^5$School of Physics and Astronomy, University of Leeds, Leeds LS2 9JT, United Kingdom\\
$^6$Department of Physics, Kyoto Sangyo University, Kyoto 603-8555, Japan\\
$^7$Instituto de Ciencia de Materiales de Madrid (ICMM), CSIC, Cantoblanco, 28049 Madrid, Spain\\
$^8$Physics Institute II, University of Cologne, 50937 K\"oln, Germany
}

\begin{abstract}
We have performed scanning angle-resolved photoemission spectroscopy with a nanometer-sized beam spot (nano-ARPES) on the cleaved surface of Pb$_5$Bi$_{24}$Se$_{41}$, which is a member of the (PbSe)$_5$(Bi$_2$Se$_3$)$_{3m}$ homologous series (PSBS) with $m$ = 4 consisting of alternate stacking of the topologically-trivial insulator PbSe bilayer and four quintuple layers (QLs) of the topological insulator Bi$_2$Se$_3$. This allows us to visualize a mosaic of topological Dirac states at a nanometer scale coming from the variable thickness of the Bi$_2$Se$_3$ nano-islands (1-3 QLs) that remain on top of the PbSe layer after cleaving the PSBS crystal, because the local band structure of topological origin changes drastically with the thickness of the Bi$_2$Se$_3$ nano-islands. A comparison of the local band structure with that in ultrathin Bi$_2$Se$_3$ films on Si(111) gives us further insights into the nature of the observed topological states. This result demonstrates that nano-ARPES is a very useful tool for characterizing topological heterostructures.
\end{abstract}

\maketitle

Three-dimensional topological insulators (TIs) are a novel quantum state of matter where gapless surface states (SSs) emerge inside the bulk band gap inverted by a strong spin-orbit coupling.\cite{1,2,3} Topological SSs are characterized by a Dirac-cone energy dispersion with a helical spin texture and play an essential role in realizing exotic quantum phenomena. The discovery of TIs has also stimulated the exploration of new topological materials beyond TIs such as topological semimetals hosting Dirac/Weyl fermions in the bulk band structure\cite{4} and topological superconductors with Majorana fermions.\cite{5,6} While bulk crystals and thin films have been the main platforms for investigating topological properties/phenomena so far, multilayer heterostructures (superlattices) consisting of alternately-stacked two or more different materials are now attracting much attention as a promising platform.\cite{7,8,9,10,11,12,13,14,15,16,17,18,19,20,21,22}

In heterostructures, one can control the thickness and stacking sequence of layers and/or insert different building blocks to modulate physical parameters (e.g., spin-orbit coupling, hybridization effect, and crystal symmetry) and obtain ordered phases (magnetism and superconductivity). This wide tunability in heterostructures is a great advantage to realize novel topological phases and gigantic quantum effects absent in parent materials. For example, for a heterostructure of TI and nontopological material where each TI interface hosts topologically-protected states (interface states; ISs), it has been proposed that precise control of the hybridization strength between adjacent ISs and/or the coupling between ISs and a certain order in the building block would lead to various novel phenomena. This is highlighted by a topological phase transition in a heterostructure of TI and nontopological insulator (NI),\cite{9} a magnetic Weyl semimetal phase in a magnetically-doped TI/NI heterostructure,\cite{7} and a Weyl superconductor phase in a magnetically-doped TI/superconductor heterostructure.\cite{8}

Since such novel properties of topological heterostructures are governed by the bulk electronic states and topological ISs/SSs that are strongly modified by the thickness, stacking sequence, or combination of building blocks, a systematic experimental determination of electronic states in each building block as a function of structural parameters is of particular importance to advance the exploration of exotic phenomena in topological heterostructures. It would thus be useful to utilize angle-resolved photoemission spectroscopy (ARPES) to directly observe the momentum-resolved band structure. However, ARPES study of topological heterostructures is rather scarce, partly because of the presence of multiple domains that usually appear on the surface. Since the size of such domains is typically much smaller than the beam spot of conventional ARPES apparatus ($\phi \sim$ 100 $\mu$m), it is very hard to obtain ARPES data by specifying each building block. To overcome such a problem, it is of particular importance to uncover the complexity of the electronic structure characterized by the entanglement of IS, SS and bulk electronic bands not only by high energy and momentum resolution but also by lateral resolution in the real space, in the scale of the nanometer.

\begin{figure*}
\includegraphics[width=5.5in]{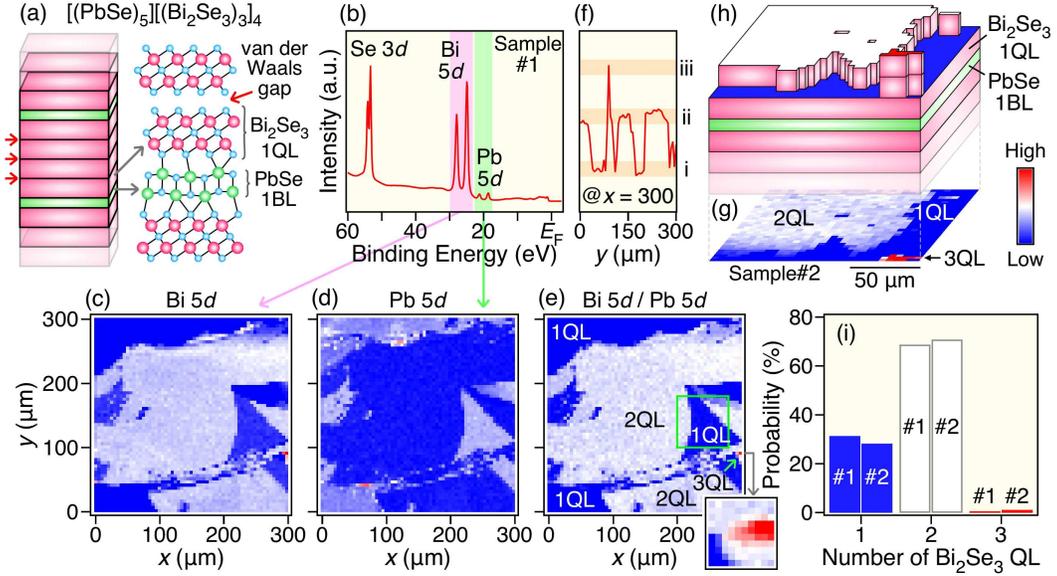}
\vspace{0cm}
\caption{(a) Schematics of the crystal structure of (PbSe)$_5$(Bi$_2$Se$_3$)$_{3m}$ (PSBS; $m$ = 4) consisting of alternately-staked one bilayer (BL) of rock-salt PbSe and four quintuple layers (QLs) of Bi$_2$Se$_3$. (b) Energy-distribution curve (EDC) in a wide energy range measured at $h\nu$ = 100 eV for sample \#1. (c, d) Intensity maps of Bi $5d$ and Pb $5d$ core levels, respectively, as a function of sample position ($x$ and $y$). The sizes of a beam spot and a scanning step were set to 120 nm and 5 $\mu$m, respectively. The intensity was obtained by integrating the spectral intensity within the energy window indicated by magenta or green shade in (b). (e) Intensity map of Bi-$5d$/Pb-$5d$ ratio. Right bottom panel shows a magnified view of 10$\times$10-$\mu$m$^2$ area which includes the 3-QL island. (f) Intensity profile at $x$ = 300 $\mu$m in (e) plotted as a function of $y$. (g) Intensity map of Bi-$5d$/Pb-$5d$ ratio obtained for sample \#2. (h) Schematic view of Bi$_2$Se$_3$ islands at cleaved surface of PSBS ($m$ = 4). (i) Histogram of the relative ratio of $n$-QL islands ($n$ = 1-3) estimated from the core-level intensity for two different cleaved surfaces (\#1 and \#2).}
\end{figure*}

In this Letter, we report the first nano-ARPES study of a topological heterostructure. The material is Pb$_5$Bi$_{24}$Se$_{41}$, a new member of naturally-occurring bulk TI/NI heterostructure (PbSe)$_5$(Bi$_2$Se$_3$)$_{3m}$ (called PSBS) with $m$ = 4,\cite{23,24,25,26} which consists of alternately-stacked four quintuple layers (QLs) of TI (Bi$_2$Se$_3$) and one bilayer of NI (PbSe) (see Figure 1a). By utilizing a very small beam spot of $\phi \sim$ 120 nm with a piezo-based sample positioning system, we have directly visualized a marked spatial variation of the valence-band structure and core-level states in PSBS and determined the size, shape, and distribution of different domain structures. Remarkably, selective probing of each domain revealed the evolution of the bulk bands and topological ISs/SSs as a function of the thickness of Bi$_2$Se$_3$ layers on a single sample. The present study not only provides important insight into the topological property of PSBS but also demonstrates the great advantage of nano-ARPES for investigating the electronic states of topological heterostructures.

High-quality single crystal of PSBS ($m$ = 4) was grown by a combination of modified Bridgman and self-flux methods using high purity elements (Pb 99.998\%, Bi and Se 99.9999\%) in a sealed evacuated quartz tube.\cite{23} Ultrathin films of Bi$_2$Se$_3$, used as a reference, were grown on Si(111) by the molecular-beam-epitaxy (MBE) method. The film thickness was controlled by the deposition time at a constant deposition rate. The actual thickness was determined by the oscillation pattern of reflection high-energy electron diffraction (RHEED). ARPES measurements for PSBS were performed with a MBS-A1 electron analyzer with energy-tunable synchrotron light at the ANTARES beamline in SOLEIL and with a Scienta-Omicron SES2002 spectrometer at the BL28A in Photon Factory (PF), KEK. We used circularly polarized light of 100 eV in SOLEIL and 40-80 eV in PF. Nano-ARPES measurements in SOLEIL were performed with submicron beam spot achieved by a Fresnel-zone-plate focusing system combined with a precise sample scanning system, optimized for an energy photon flux of 100 eV.\cite{27,28} PSBS crystals were cleaved in-situ in an ultrahigh vacuum better than 1$\times$10$^{-10}$ Torr along the (111) crystal plane. ARPES measurements for Bi$_2$Se$_3$ thin films were carried out with an ARPES-MBE combined system at Tohoku University equipped with a MBS-A1 electron analyzer. We used the He-I$\alpha$ ($h\nu$ = 21.218 eV) line to excite photoelectrons. PSBS crystals and Bi$_2$Se$_3$ thin films were kept at $T$ = 60 K in SOLEIL and 30 K in Tohoku University during ARPES measurements.

\begin{figure}
\includegraphics[width=3in]{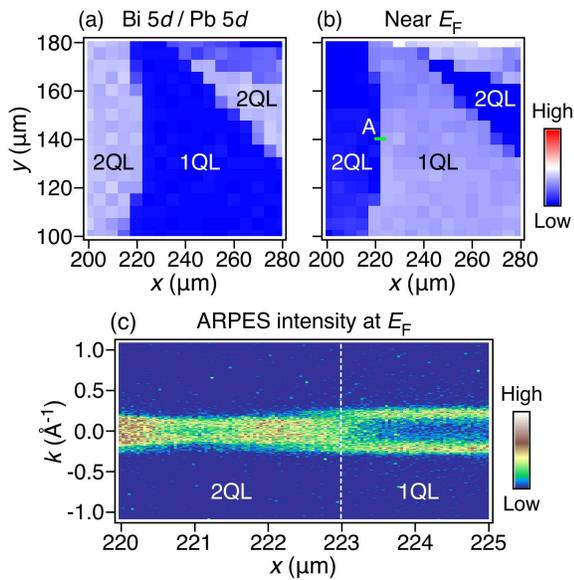}
\vspace{0cm}
\caption{(a) Magnified view of the region enclosed by green square in Figure 1(e). (b) Intensity map near $E_{\rm F}$ measured in the same surface area with (a) [green square in Figure 1(e)]. Intensity was obtained by integrating the spectral intensity within 0.75 eV below $E_{\rm F}$. (c) Spatial variation of the momentum-dependent ARPES intensity in the range of $\pm$50 meV centered at $E_{\rm F}$ measured along green line (cut A) in (b); the momentum $k$ is taken along the $k_y$ direction of the Brillouin zone.
}
\end{figure}

First, we performed a scanning core-level photoemission microscopy on PSBS ($m = 4$) surface with a nano-focused beam. A representative energy-distribution curve (EDC) at a fixed sample position (Figure 1b) shows several shallow core levels of Se $3d$, Bi $5d$, and Pb $5d$. Figures 1c,d display real-space maps of Bi-$5d$ and Pb-$5d$ core-level intensities, respectively, measured on an area of $300 \times 300$ $\mu$m$^2$. One can recognize a mosaic-like pattern that mainly consists of blue- and white-colored regions separated by a sharp boundary, indicative of formation of multiple domain structures on the cleaved surface. Their spatial distribution is random and the area of each domain varies from $\sim$200$\times$200 $\mu$m$^2$ to $\sim$5$\times$5 $\mu$m$^2$, the latter of which is too small to resolve by conventional ARPES with a beam spot of $\sim$100 $\mu$m. It is also noted that blue-colored (white-colored) domain in Figure 1c shows white (blue) color in Figure 1d. Because of this anticorrelation between the Bi-$5d$ and Pb-$5d$ core-level intensities, an enhanced color contrast is obtained by taking their relative intensity (see Figure 1e). In Figure 1e, besides blue- and white-colored domains, a small but finite red-colored domain ($\sim$4$\times$7-$\mu$m$^2$ area) is observed at ($x$, $y$) $\sim$ (300 $\mu$m, 100 $\mu$m) (indicated by a green arrow), as can be seen more clearly in a fine map of 10$\times$10-$\mu$m$^2$ area which includes the red-colored region (see right bottom panel of Figure 1e). The existence of three types of domains is confirmed by a line profile at $x = 300$ $\mu$m (see Figure 1f) where the Bi-$5d$/Pb-$5d$ intensity falls into three levels i-iii that correspond to blue-, white-, and red-colored domains in Figure 1e, respectively, and is further corroborated by the observation of a tricolor spatial image on a different sample surface (Figure 1g). The three-domain nature of PSBS ($m$ = 4) surface is naturally understood in terms of three different cleavage planes in the crystal. Namely, the cleavage may preferably occur at each van der Waals gap between QLs within the Bi$_2$Se$_3$ unit (red arrows in Figure 1a), leaving 1-3 QLs of Bi$_2$Se$_3$ nano-islands on the PbSe layer. Specifically, blue-, white-, and red-colored regions in Figures 1e,g would correspond to 1-, 2-, and 3-QL Bi$_2$Se$_3$ islands, respectively (see schematic in Figure 1h), because the finite photoelectron escape depth leads to the stronger Bi-$5d$/Pb-$5d$ intensity ratio for thicker Bi$_2$Se$_3$ domains. It is noted that the PbSe layer and the neighboring Bi$_2$Se$_3$ QL are connected with each other via Pb-Se bonds\cite{29} (see Figure 1a), so that cleavage between these two layers would rarely occur in PSBS, except for small $m$ phases (e.g., $m$ = 1). Our atomic-force microscopy measurements on PSBS with $m$ = 4 revealed a step structure with $\sim$1-nm height that is close to the thickness of 1-QL Bi$_2$Se$_3$, supporting the preferable cleavage between Bi$_2$Se$_3$ QLs (see Figure S1 in Supporting Information). Figure 1i displays the probability to observe each domain, estimated from the maps in Figures 1e,g (sample \#1 and \#2, respectively). A low probability (1-3 \%) for seeing the 3-QL domain (also see Figures 1e,g) highlights the crucial importance of a real-space imaging with nano-focused beam spot to resolve all three domains. It is noted that the multiple domain structure was overlooked in a previous ARPES study on PSBS ($m$ = 1 and 2) with a large beam spot.\cite{14}

The selective observation of 1-3 QLs of Bi$_2$Se$_3$ islands by nano-ARPES provides an excellent opportunity to simultaneously investigate the band structure of Bi$_2$Se$_3$ with various thicknesses in one sample. Figures 2a,b display a comparison between the spatially resolved core-level and near-$E_{\rm F}$ intensities measured in the same region of cleaved surface (green square in Figure 1e). One can identify similar mosaic pattern with reversed contrast between Figures 2a and 2b, such as a vertical domain boundary at $x$ $\sim$ 220 $\mu$m and a diagonal domain boundary at upper right. This indicates a modification of the local band structure depending on the number of QLs. By using momentum-resolving capability of nano-ARPES, we further visualized the evolution of Fermi surface across the boundary between the 1-QL and 2-QL domains (along a cut A in Figure 2b). The ARPES intensity at $E_{\rm F}$ as a function of $k$ and $x$ (Figure 2c) shows the presence of a single Fermi surface with the Fermi wave vector of $\sim$0.2 ${\rm \AA}^{-1}$ in the 1-QL domain, and a sudden change in the Fermi vector by passing through the domain boundary into the 2-QL domain, as represented by appearance of strong intensity at $k \leq$ 0.1 ${\rm \AA}^{-1}$. The observed domain-dependent near-$E_{\rm F}$ intensity and Fermi surface are related to the in-plane mosaic of topological SSs/ISs as discussed later.

To gain further insight into the local band structure of Bi$_2$Se$_3$ nano-islands, we have performed high energy- and momentum-resolution ARPES measurements on each domain. In Figure 3a, a single parabolic band is observed for the 1-QL domain, consistent with a single Fermi surface observed in Figure 2c. On the other hand, multiple bands are resolved for the 2-QL domain, and the number of bands is further increased in the 3-QL domain. The observed domain-dependent change in the local band structure would be mainly triggered by quantum confinement within the Bi$_2$Se$_3$ layer.\cite{14} To understand the origin of the band dispersions in each domain, it is useful to compare the present ARPES results with those for ultrathin Bi$_2$Se$_3$ films, in which the band structure is strongly reconstructed by the electron confinement effect.\cite{30} As seen from Figure 3b, the overall band structure of Bi$_2$Se$_3$ ultrathin films on Si(111) substrate (Bi$_2$Se$_3$/Si) resembles that of PSBS. For example, 1-QL Bi$_2$Se$_3$/Si has a single parabola as in the 1-QL domain of PSBS, and 2-QL Bi$_2$Se$_3$/Si has a bandgap at $E_{\rm B}$ $\sim$ 0.6 eV with a ``M"-shaped dispersion of the lower hole-like branch as in the 2-QL domain of PSBS. This similarity strongly suggests that a parabolic band at the $\bar{\Gamma}$ point in the 1-QL domain of PSBS (black dots in left panel of Figure 3a) is ascribed to the quantized bulk conduction band (CB), as is the case of Bi$_2$Se$_3$/Si.\cite{30} Also, for the 2-QL domain, two inner electron pockets (black dots) are quantized bulk CBs with a Rashba-type splitting, while two outer electron pockets and hole-like dispersions at higher $E_{\rm B}$'s (red dots) are the upper and lower branches of gapped Dirac cones that originate from the hybridized topological SSs/ISs (note that the Rashba splittings of CB and hybridized topological SSs/ISs are a consequence of the inversion-symmetry breaking at the surface as observed in Bi$_2$Se$_3$/graphene\cite{30}). Upon increasing the film thickness to 3 QL, the bulk valence band (VB) is well isolated from the lower Dirac band (see black dots at $E_{\rm B}$ $\sim$ 0.7 eV in right panel of Figure 3b), accompanied by a reduction of hybridization gap at the Dirac point (see red dots). By taking into account these trends, the energy bands that produce the minimum band gap in the 3-QL domain of PSBS (red dots in right panel of Figure 3a) are assigned to the hybridized topological SSs/ISs, while other bands (black dots) are attributed to the quantized bulk bands; i.e., electron-like bands near $E_{\rm F}$ and hole-like bands at higher $E_{\rm B}$'s are the quantum well states of CB and VB, respectively.

\begin{figure}
\includegraphics[width=3in]{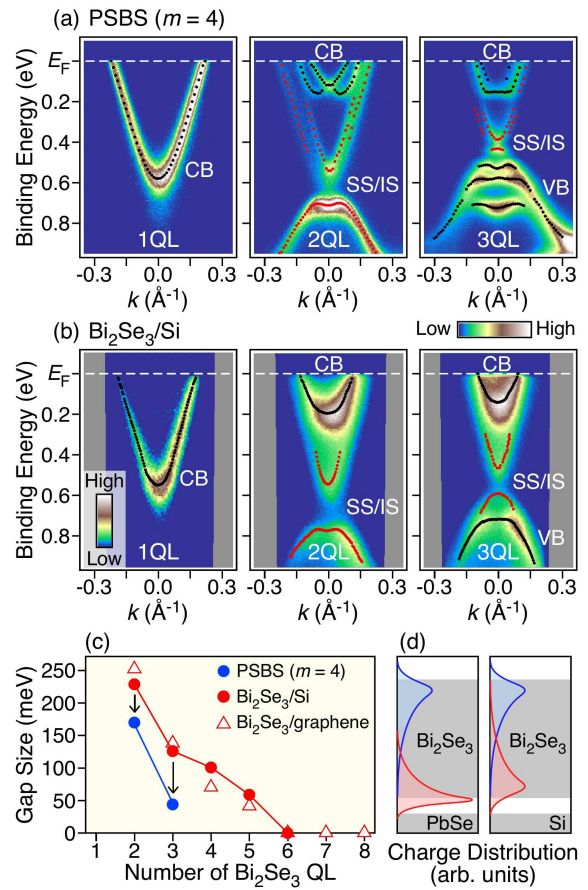}
\vspace{0cm}
\caption{(a) ARPES intensity plots to show the near-$E_{\rm F}$ band dispersions around the $\bar{\Gamma}$ point for $n$-QL Bi$_2$Se$_3$ islands ($n$ = 1-3); the momentum $k$ is taken along the $k_y$ direction of the Brillouin zone. Dots show experimental band dispersions determined by tracing the peak position of EDCs. (b) Near-$E_{\rm F}$ band dispersions of $n$-QL Bi$_2$Se$_3$ ultrathin films ($n$ = 1-3) grown on Si(111). (c) The size of the energy gap at the Dirac point as a function of the number of QLs ($n$) for PSBS, Bi$_2$Se$_3$/Si, and Bi$_2$Se$_3$/graphene\cite{30} estimated from the numerical fittings of EDCs at the $\bar{\Gamma}$ point. (d) Schematic illustration of the spatial distribution of the topological surface and interface states in (left) PSBS and (right) Bi$_2$Se$_3$/Si.}
\end{figure}

The present systematic investigation of the near-$E_{\rm F}$ electronic states by domain-selective nano-ARPES allows us to discuss the evolution of topological properties as a function of the Bi$_2$Se$_3$ thickness (1-3 QLs). As seen in Figure 3a, while an inverted band structure in the 2- and 3-QL domains of PSBS leads to the appearance of topological SSs and ISs, their mutual hybridization results in gap-opening at the Dirac point. The observed hybridization-gap size is 170 meV and 40 meV for the 2- and 3-QL domains, respectively. Intriguingly, these values (see blue circles in Figure 3c) are substantially smaller than those for ultrathin Bi$_2$Se$_3$ films on various substrates such as Bi$_2$Se$_3$/Si and Bi$_2$Se$_3$/graphene\cite{30} (red circles and triangles, respectively). For example, the hybridization gap in the 3-QL domain of PSBS is as small as that in 5-QL Bi$_2$Se$_3$/Si and Bi$_2$Se$_3$/graphene. This suggests that the critical thickness of Bi$_2$Se$_3$ layer at which a gapless topological phase emerges in PSBS would be smaller than 6 QL that has been proposed for Bi$_2$Se$_3$ films.\cite{30} Possibly, the gapless topological phase may be realized in bulk PSBS with $m$ = 4. Such robustness of the topological phase down to ultrathin regime may be potentially useful for realizations of gigantic topological effects in bulk crystals as well as for future application to nano-scale devices. It is noted that high-resolution ARPES measurements on a different sample confirmed that the observed small hybridization gap is an intrinsic property of PSBS (see Figure S2 in Supporting Information).

The origin of the small hybridization gap in PSBS may be ascribed to a difference in the spatial distribution of topological ISs between PSBS and Bi$_2$Se$_3$/Si. It has been predicted for TI/NI heterostructures that the interaction with an interfaced NI may strongly modify the real-space distribution of topological ISs; for instance, topological ISs may migrate from TI to interfaced NI.\cite{9} Therefore, one may expect that the wave function of topological ISs is localized at the interface or even shifted toward the PbSe layer compared with that in Bi$_2$Se$_3$/Si (see red shades in the schematics of Figure 3d) and the resultant reduction of the overlap between the topological SSs within the same Bi$_2$Se$_3$ unit would suppress the hybridization gap at the Dirac point, as observed in the present study. A possible origin of such a change in the spatial distribution of topological ISs can be the in-plane tensile strain introduced in the Bi$_2$Se$_3$ layer of PSBS through the lattice mismatch with the PbSe layer\cite{17,23} (note that Bi$_2$Se$_3$/Si and Bi$_2$Se$_3$/graphene are almost strain-free). This scenario is supported by previous first-principles calculations which demonstrated that tensile strain reduces the hybridization gap of topological SSs/ISs in ultrathin Bi$_2$Se$_3$ layers.\cite{31} Moreover, recent first-principles calculations for PSBS ($m$ = 2) with realistic lattice parameters, in which the strain effect is naturally taken into account, show a substantially small hybridization gap of $\sim$20 meV (c.f., 100-200 meV for unstrained 2-QL Bi$_2$Se$_3$).\cite{32} It is noted that, while strain effect may also arise from crystal imperfections, the observed sample insensitivity of the hybridization gap size (Figure S2) rules out such a possibility because the crystal imperfection, if it exists, would be sample-dependent and hence the gap size should change from sample to sample. In addition, surface reconstruction that may also lead to lattice strain at the surface is also ruled out because our band-structure mapping over a wide momentum space does not show a signature of band-folding indicative of surface reconstruction.

In summary, we have visualized the spatial variation of the core-level intensity in PSBS ($m$ = 4) by nano-ARPES measurements, and uncovered the three-domain structure formed with 1-3 QLs of Bi$_2$Se$_3$ nano-islands on PbSe layer. We observed the domain-dependent drastic changes in the topological Dirac-cone dispersion. Comparison with the band dispersions of Bi$_2$Se$_3$/Si revealed a substantial reduction of the hybridization gap between topological SSs and ISs in PSBS, which indicates that the PbSe layer is a useful building block that helps the Bi$_2$Se$_3$ layer to maintain its topological character. The present result demonstrates that nano-ARPES is a powerful method to investigate the bulk and topological states of nano-islands with different thicknesses and is thus useful for unraveling the peculiar properties of topological heterostructures.

\begin{acknowledgments}
We thank H. Kimizuka, K. Horiba, and H. Kumigashira for their assistance in the ARPES measurements and thin-film growth. We also thank KEK-PF for beamline BL28A (Proposal number: 2018S2-001), and SOLEIL for beamline ANTARES (Proposal number: 20171255). This work was supported by JST-PRESTO (No: JPMJPR18L7), JST-CREST (No: JPMJCR18T1), MEXT of Japan (Innovative Area ``Topological Materials Science" JP15H05853, JP15K21717), and JSPS (JSPS KAKENHI No: JP17H01139, JP26287071, JP25220708, JP18J20058, and JP18H01160). The work in Cologne was funded by the Deutsche Forschungsgemeinschaft (DFG, German Research Foundation) - Project number 277146847 - CRC 1238 (Subproject  A04).
\end{acknowledgments}

\bibliographystyle{prsty}

\end{document}